\documentclass[12pt,onecolumn] {article}
\usepackage{amsmath}
\usepackage{amssymb}
\usepackage{latexsym}
\usepackage{graphicx}

\begin{document}

\begin{center}
\baselineskip=24pt

{\Large \bf Muon simulation codes MUSIC and MUSUN for underground physics}

\baselineskip=18pt

\vspace{0.3cm}
V.~A.~Kudryavtsev \footnote{Corresponding author; address: 
Department of Physics and Astronomy, University of Sheffield,
Sheffield S3 7RH, UK, e-mail: v.kudryavtsev@sheffield.ac.uk},

\vspace{0.3cm}
{\it Department of Physics and Astronomy, University of Sheffield, Sheffield S3 7RH, UK}\\

\vspace{0.3cm}
\begin{abstract}
The paper describes two Monte Carlo codes dedicated to muon simulations: MUSIC 
(MUon SImulation Code) and MUSUN (MUon Simulations UNderground). MUSIC
is a package for muon transport through matter. It is particularly useful for propagating
muons through large thickness of rock or water, for instance from the surface down to
underground/underwater laboratory. MUSUN is designed to use the results of muon
transport through rock/water to generate muons in or around underground laboratory
taking into account their energy spectrum and angular distribution.
\end{abstract}

\end{center}



\pagebreak

\section {Introduction}
\label{intro}

Muon transport through matter plays an important role in many areas of particle 
and astroparticle physics. Cosmic-ray muons are detected at large depths
underground and underwater (here and hereafter we use the term underwater that
includes also under-ice experiments). They are used to study the energy spectrum and
composition of primary cosmic rays and calculations of their fluxes, energy
and angular distributions are the key element of this research (see, for instance,
Refs. \cite{gaisser,macro1,lvd1,lvd2,macro}. 

Experiments with high-energy muon neutrino beams from accelerators require
accurate simulations of muon transport from the point of neutrino interaction to the
detector. Similarly, neutrino telescopes are detecting (or expecting to detect) muons
from atmospheric and astrophysical neutrinos, and three-dimensional propagation
of muons from their production point to the detector is crucial for the interpretation of
experimental data \cite{baikal,amanda,antares}.

Cosmic-ray muons are also a background in experiments looking for rare events at low
and high energies deep underground or underwater. Atmospheric down-going muons 
can be erroneously reconstructed
as upward-going muons that mimic neutrino-induced events in a search for astrophysical
neutrinos at GeV-TeV energies 
or in an atmospheric neutrino detection for neutrino oscillation studies.
Cosmic-ray muons also produce secondary neutrons (with MeV-GeV energies) by 
interacting with rock. These neutrons can mimic low-energy (keV-MeV) events 
in detectors looking for WIMP (Weakly Interacting Massive Particle) dark matter,
neutrinoless double-beta decay and neutrinos (solar, geophysical, supernova 
neutrinos, etc.) (see Ref. \cite{formaggio} for a review and 
Refs. \cite{wang,musun,carson04,araujo05} 
for example calculations of muon-induced neutron
fluxes underground) .
High-energy (GeV) neutrons from muons can produce events with
a signature similar to proton decay.

There are a few more applications from different areas of science. A morphological 
reconstruction of mountains and natural caves 
using atmospheric muons was suggested in Ref. \cite{caffau}.
A search for hidden chambers in pyramids was discussed in Ref. \cite{alvarez}.
A `muon radiography' using multiple scattering of cosmic-ray muons
was proposed recently \cite{schultz} to discriminate between low-A and high-A
materials in cargo.

All applications mentioned above require accurate calculations of muon spectra and 
scattering beyond a slab of material. Most of them involve muon transport through
large thickness of matter. Hence the CPU time should be 
reduced to a minimum without compromising the accuracy of calculations.

Several Monte Carlo codes are able to transport muons through matter with high
accuracy. The codes can be split in two categories: (i) multipurpose
particle transport codes, such as GEANT4 \cite{geant4} and FLUKA \cite{fluka}, 
and (ii) codes developed specifically for muon propagation through large thickness of 
material, such as PROPMU \cite{propmu}, MUSIC \cite{music,music1}, MUM \cite{mum} 
and MMC \cite{mmc}. 

Significant progress has recently been achieved in the development of the
multipurpose transport codes for particle physics applications. The codes have become 
faster, more robust, flexible and accurate. However, their flexibility requires a good
knowledge of physics and programming skills from a user. GEANT4, for instance,
is designed as a powerful toolkit but a good knowledge of the code including
models and programming language is needed to use it properly. Significant
efforts and time are required to become familiar with such a toolkit. These codes
are absolutely necessary when simulating events consisting of many particles that should be
produced, transported and detected practically at the same time.
Meanwhile, some tasks, for instance muon transport through a homogeneous material, 
may be accomplished 
without using multipurpose codes. If a user is interested in transporting muons 
without following secondary particles produced by them, it is enough to consider 
accurately only muon interactions and muon energy losses 
neglecting the fate of secondaries. This is the idea implemented in specially
developed muon transport codes.

In this paper we describe the three-dimensional muon propagation code MUSIC. 
Although the first version of the code
was released in 1997 \cite{music} and several modification were reported since 
then \cite{music1}, 
we believe that the recent developments and improvements
made to the code and the variety of applications should be described in a separate
paper. The second part of the paper is dedicated to the code 
MUSUN 
written to sample muons underground \cite{musun} or underwater \cite{musun1}
using the results of
muon transport carried out with MUSIC.

\section {Muon transport through large thickness of matter: MUSIC}
\label{music}

The first version of MUSIC (MUon SImulation Code), written in FORTRAN, 
has been released in 1997
\cite{music}. It has been used in the interpretation of data from the LVD experiment
in the Gran Sasso Laboratory, namely in the reconstruction of the
muon energy spectrum at surface from the measured depth -- vertical muon 
intensity relation \cite{lvd1}
and in the evaluation of the fraction of prompt muons in the high-energy muon flux
\cite{lvd2}. Several improvements have been done to the code and new features 
have become available since then.
The basic features of the code and improvements are described below. 

The code takes into account
the energy losses of muons due to four processes: ionisation (using Bethe-Bloch formula)
including knock-on electron production, bremsstrahlung (or braking radiation),
electron-positron pair production and muon-nucleus 
inelastic scattering (or photonuclear interactions).
The cross-section of bremsstrahlung was taken from Ref. \cite{kkp} in the first version of MUSIC
with an option to use the cross-section from Refs. \cite{bb1,ab1}.
The effect of different cross-sections on the MUSIC results was studied in Ref. \cite{music}.
The correction to the Born approximation (Coulomb correction) 
for bremsstrahlung cross-section was not taken into account. It was shown
\cite{ab1} that this correction does not exceed 1\% even for heavy nuclei.
The pair production cross-section was taken from Ref. \cite{kp} in the first version
of MUSIC.
New parameterisation of the pair production on atomic electrons \cite{kelner}
has been implemented in the second version of the code \cite{music1}.
Original parameterisation of muon inelastic scattering cross-section \cite{bb}
has been complemented by a more accurate treatment suggested in Ref. \cite{bs}.
An option has been added to calculate this cross-section using the deep-inelastic 
scattering formalism and nucleon structure functions suggested in Ref. \cite{dutta}.
The ALLM parameterisations \cite{allm} of structure functions have been implemented.
The default options in MUSIC (the recommended cross-sections) are:
(i) bremsstrahlung -- Ref. \cite{kkp}; (ii) pair production -- Refs. \cite{kp,kelner}; 
(iii) inelastic scattering -- Ref. \cite{bs}. All results presented here have been obtained with 
this set of cross-sections.

Muon interaction cross-sections are calculated in MUSIC at the beginning of the first run (or
by the code developer) for all elements present in a material and specified by a user,
and are averaged using the weights (a fraction of each element by mass) provided
by the user.

There are two major versions of MUSIC existing and developed in parallel: 
(i) `standard', dedicated for muon transport
through large thickness of matter; (ii) `thin slab', developed for muon transport through
thin slabs of materials.

The standard version of MUSIC considers all interaction processes stochastically
if the fraction of energy lost by a muon 
in the interaction exceeds a pre-defined value of a parameter, 
$v_{cut}$ (see Ref. \cite{music} for the full code description and tests). 
The value of $v_{cut}$ can vary from $10^{-5}$ to 1 and 
can be set by a user but the recommended value taken as a compromise between
the accuracy of the code and its speed, is $10^{-3}$ as in the first
version of the code \cite{music}. The program evaluates the mean free path of a muon
between two subsequent interactions (with $v > v_{cut}$) 
using the sum of integrated cross-sections,
where integrals are computed between $v_{cut}$ and 1. Then it samples the real
path of the muon to the next interaction using a random number generator from
the CERN library (RANLUX). Another random number determines the type of the
interaction. The third random number is used to select the fraction of muon energy
lost in the interaction $v$. Then the code calculates the continuous energy losses
of the muon between the two interactions, 
i.e. mean energy losses due to the four aforementioned processes
with $v < v_{cut}$. Mean energy loss due to ionisation
is computed using Bethe-Bloch formula for $v < v_{cut}$. Knock-on
electron production is added to stochastic processes
at $v \ge v_{cut}$ including corrections to the ionisation
energy loss due to $e$-diagrams for muon bremsstrahlung on an electron suggested
in Ref. \cite{kkp}. In this process the photon is emitted by the electron and is 
accompanied by the high-energy recoiling electron.

Muon deflection due to multiple Coulomb scattering in the plane perpendicular 
to the initial muon direction is calculated
between every two interactions with $v \ge v_{cut}$ \cite{music}. 
The process is treated in the Gaussian approximation \cite{rossi} that was also used
in Ref. \cite{propmu}. More accurate treatment of muon angular deviation in the framework
of Moli\`ere theory results in a similar distribution of muon scattering angles beyond
the large thickness of rock with a small increase of the mean deflection angle
\cite{music}. However, the original Moli\`ere theory was developed for angular deviation 
only and does not provide the lateral displacement that is sometimes more important from
the experimental point of view (for instance, for muon bundles underground).

Although multiple Coulomb scattering dominates over stochastic processes in the muon
deflection \cite{music}, muon deviation due to other interactions is also taken 
into account in MUSIC. The angular deviation due to muon inelastic scattering is 
computed using double-differential cross-section \cite{bb}. The muon scattering angle
due to bremsstrahlung and pair production is calculated following the parameterisations
suggested in Ref. \cite{vg} (see also Ref. \cite{music} for detailed description).

Note that the muon transport code MUM \cite{mum} is
one-dimensional and does not take into account muon deflection in the plane
perpendicular to the initial direction. The code MMC \cite{mmc} considers only muon
deviation due to multiple scattering and not many details or results 
are given in the original paper \cite{mmc}. PROPMU \cite{propmu} also
treats only multiple scattering. Muon transport codes MUSIC, MUM and MMC
were found to agree with each other giving similar muon energy distributions beyond large
thickness of rock or water and similar muon survival probabilities
\cite{mum}. MUSIC and PROPMU are in agreement for
muon transport in rock \cite{music} but results obtained with PROPMU in water were 
found to be different from those obtained with MUSIC and MUM \cite{mum}.

The MUSIC code allows the transport of muons with energies up to $10^7$~GeV. All
muons are considered to be ultrarelativistic. If the total muon energy becomes smaller than
the muon mass, the muon is considered to be stopped. 
The code consists of two files and is arranged as consecutive calls to two or three 
subroutines written in FORTRAN from the `main' user program. The call to the first 
subroutine is optional: it allows calculation of the muon cross-sections and energy losses
and can be done at the beginning of the first run. The files with cross-sections and energy
losses can also be supplied by the author. The call to the second subroutine allows
reading the muon cross-sections from the computer disk. A single call to the third subroutine
transports one muon with given initial parameters (energy, coordinates and direction 
cosines) to a specified distance in a material with previously calculated cross-sections.
The code (the third subroutine) returns the muon parameters at the end of the muon
path in the material. If the muon has stopped before reaching the end of the material,
the zero value for the muon energy is returned together with approximate coordinates
of the point where the muon has stopped.

A special version of MUSIC (`thin slab')
has been developed for muon transport through
thin slabs of materials. Originally it has been written for water as part
of the software for the ANTARES experiment \cite{antares}. 
This version has been aimed
at providing muon energy, position and direction at the end
of every small segment of muon path in water and at passing
the muon energy loss at the segment to another part of software
that generated Cherenkov photons.
Since then this version has also been used to estimate
muon deflection in high-A materials (iron, lead and
uranium) for possible security applications (searching
for hidden high-A materials, like uranium, in cargo) \cite{pradip}.
The `standard' version of MUSIC, in the absence of
stochastic interactions on the small segment, always returns
the mean value for continuous energy loss at the end of the
segment. In the `thin slab' version the cut that separates 
stochastic and continuous parts of the energy loss is reduced to
1 MeV, meaning that practically all muon interactions are
stochastic. The ionisation energy loss is calculated using
Landau distribution (call to a function from the CERN library).
This version of the code allows the transport of muons with energies
up to $10^{9}$~GeV but without taking into account the LPM effect.
Both versions of MUSIC (standard and thin slab)
give consistent results for thick slabs of matter but the `thin slab'
version is more CPU consuming because of the lower value of $v_{cut}$.
The thin slab version gives more accurate results for energy spectra and
angular deviation beyond thin slabs of material.

Although the energy loss due to muon pair production by
muons is not included in the code because of its small value compared
to the electron-positron pair production, there is a possibility to
generate muon pairs along the muon path \cite{music1}.

\begin{figure}[htb]
   \includegraphics[width=15.cm]{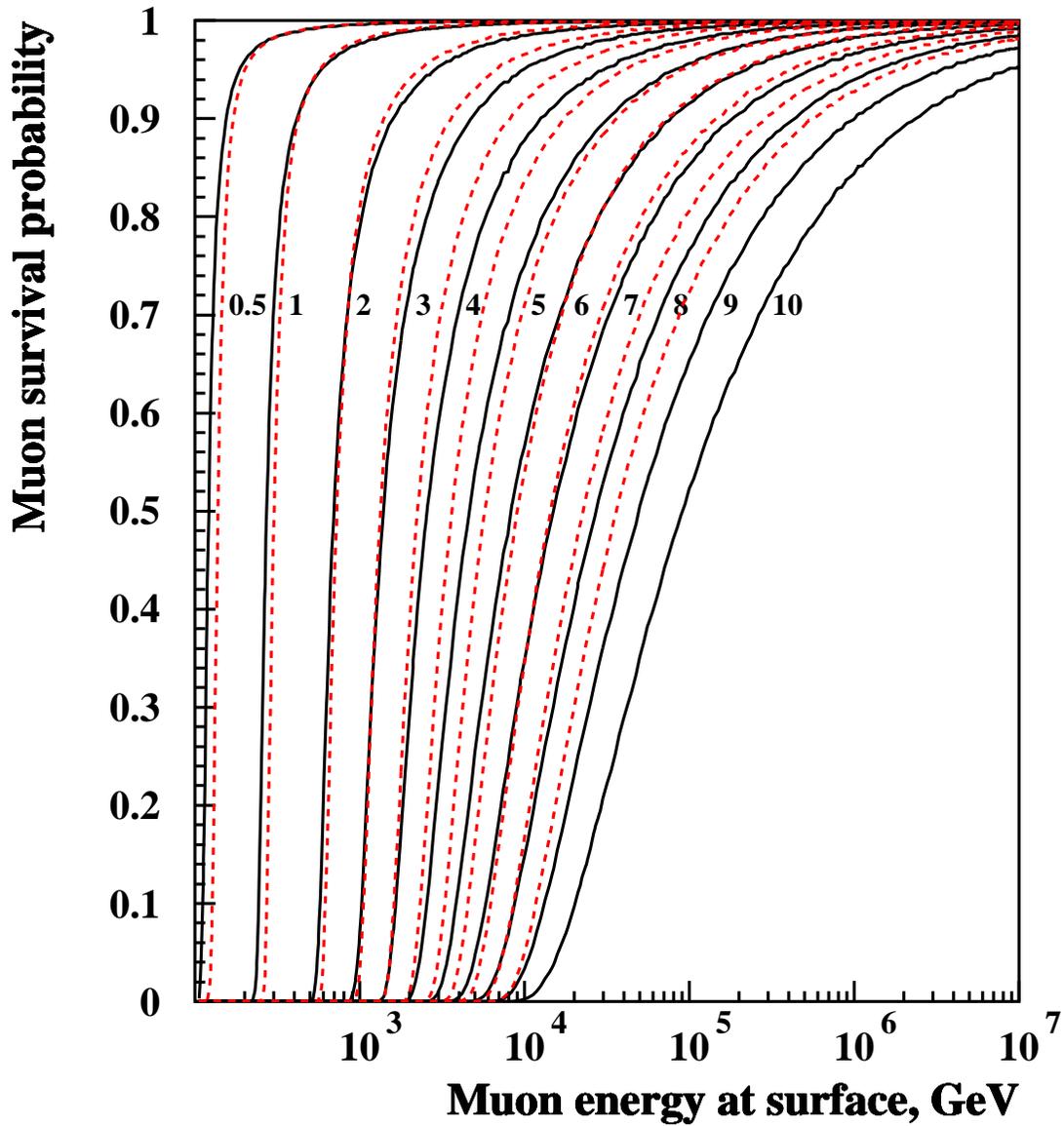}
    \caption{Survival probabilities as functions of muon energy
at surface for different depths (from 0.5 to 10 km~w.~e.) 
in standard rock (black solid curves) and water 
(red dashed curves). Numbers to the right from each solid curve show the depths in 
km~w.~e. for
standard rock. Survival probability curves for water are shifted to the right
(for small depths) or to the left (for depths larger than 1 km~w.~e.) relative to those
for standard rock.}
  \label{fig-prob}
\end{figure}

\begin{figure}[htb]
   \includegraphics[width=15.cm]{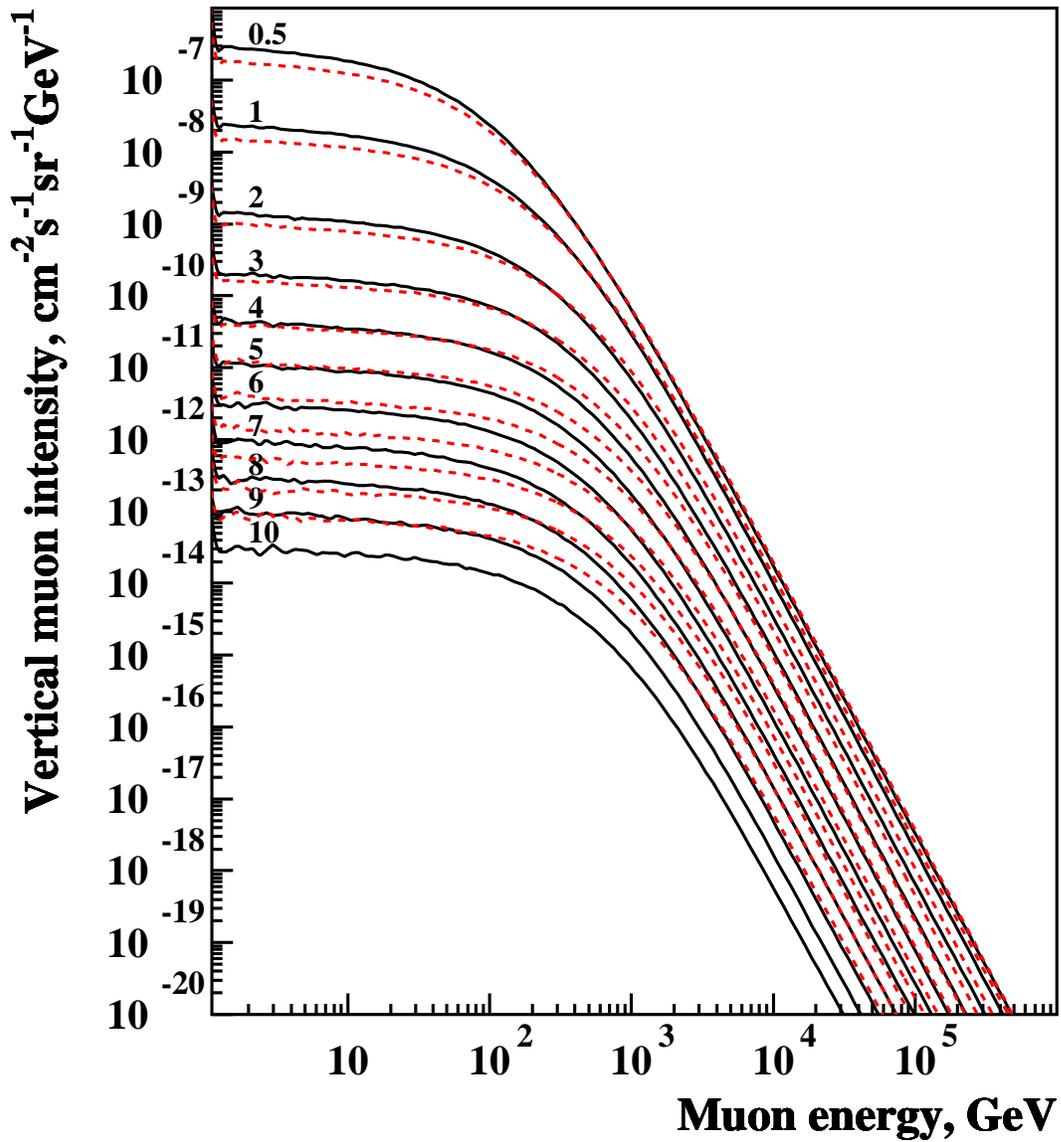}
    \caption{Muon energy spectra at vertical at different depths in standard rock (black solid curves)
and water (red dashed curves).
Numbers above the curves for standard rock show the depth in km~w.~e.
Curves for water are shifted above or below corresponding curves for standard rock.}
  \label{fig-musp-energy}
\end{figure}

\begin{figure}[htb]
   \includegraphics[width=15.cm]{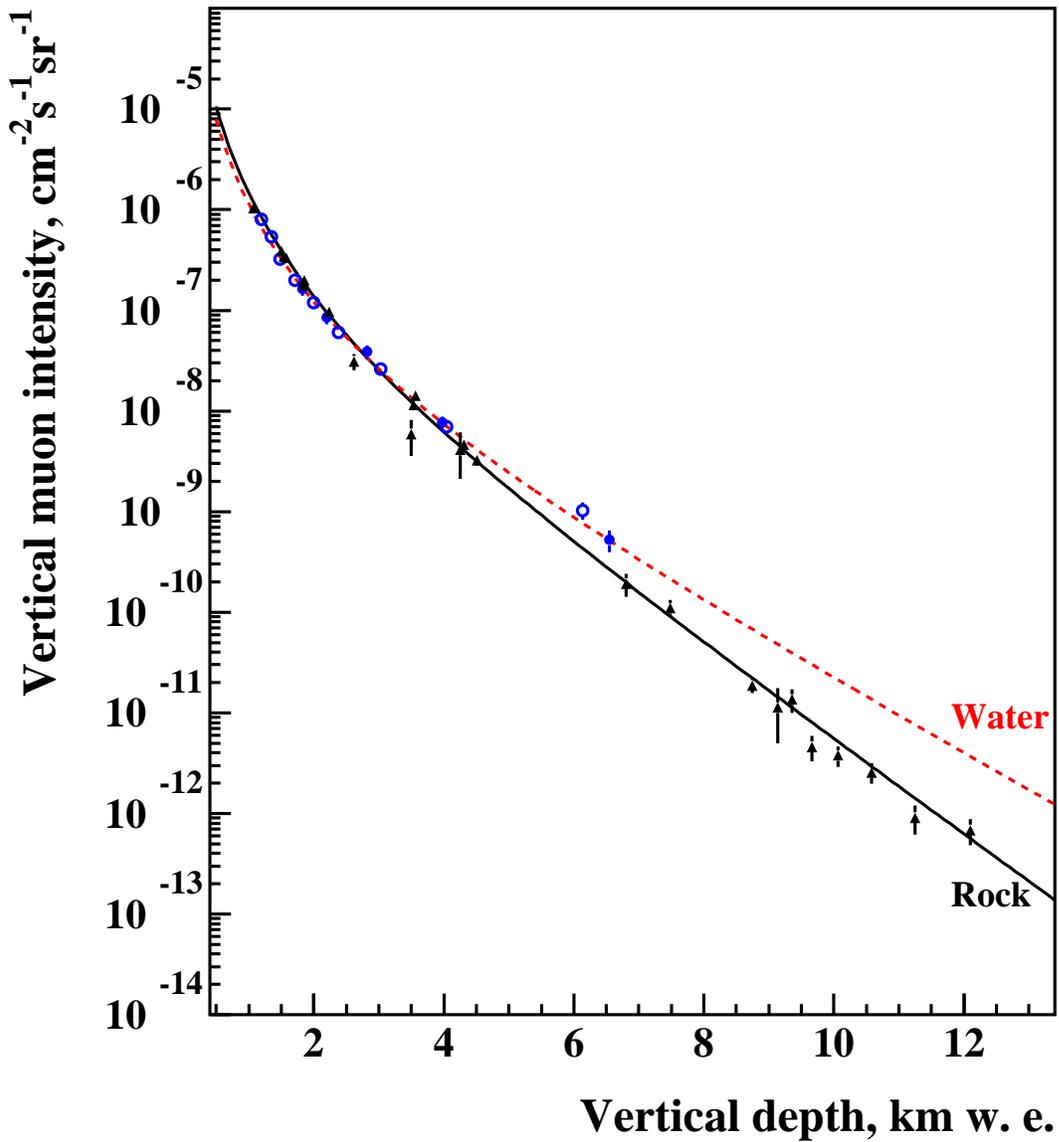}
    \caption{Depth -- vertical muon intensity relation (muon intensity at vertical
as a function of depth) for standard rock and water. The data points for standard rock 
(black triangles) are
from the compilation of experimental results \cite{crouch}. The data points for water
are from the Baikal \cite{baikal} (blue open circles) and 
AMANDA \cite{amanda} (blue filled circles) experiments.}
  \label{fig-int}
\end{figure}

A few results from muon transport through standard rock (Z=11, A=22, 
$\rho=2.65$~g/cm$^3$) and pure water are shown in Figures 
\ref{fig-prob}-\ref{fig-int}. 
Muons with initial energies ranging from $10^2$~GeV to $10^7$~GeV
were transported through 15 km~w.~e. of standard rock and water 
and their energies at different depths (distances from initial point)
were recorded. $10^5$ muons were propagated for each value of initial energy.
This means that for a survival probability (defined as a probability
for a muon with a certain initial energy to traverse a pre-defined distance) 
of 0.01, the statistical error is about 3\%. Survival probabilities as functions of muon energy
at surface for different depths are presented in 
Figure \ref{fig-prob} for standard rock (black solid curves) and water 
(red dashed curves). Numbers to the right from each solid curve show the depths 
in km~w.~e. for
standard rock. Survival probability curves for water are shifted to the right
(for small depths) or to the left (for depths larger than 2 km~w.~e.) relative to those
for standard rock. This behaviour is due to the presence of hydrogen in water.
Hydrogen has the ratio of $Z/A\approx1$ whereas most other materials have $Z/A \le 0.5$.
Ionisation energy loss is proportional to $Z/A$ whereas energy losses due
to pair production and bremsstrahlung are approximately proportional to $Z(Z+1)/A$.
At low muon energies (below 1 TeV) and small depths ionisation energy loss 
dominates over other processes, muon energy losses in water are bigger
than in standard rock  and muon survival probability for a fixed
initial energy is smaller in water than in standard rock. At intermediate depths
(between 1 and 3 km~w.~e.) the muon survival probabilities in water are smaller
for low energies and bigger for higher energies compared to standard rock.
At large depths (high muon energies) 
the energy loss due to pair production and bremsstrahlung dominate over 
ionisation and the muon survival probabilities in water are bigger than those
in standard rock for all energies.

Muon energy spectra at vertical at different depths in standard rock (black solid curves)
and water (red dashed curves)
are presented in Figure \ref{fig-musp-energy}.
Numbers above the curves for standard rock show the depth in km~w.~e.
The spectra have been calculated by convoluting muon energy distributions
underground obtained with MUSIC, with muon energy spectra at 
surface taken in the form that
fits the data from the LVD experiment \cite{lvd1,lvd2} (for full description
of the procedure see Section \ref{musun} below).
Curves for water are shifted above or below those for standard rock due to the
presence of hydrogen in water (see discussion above).

MUSIC has been extensively tested against experimental data.
It has first been used in the analysis of muon intensities measured
by the LVD experiment \cite{lvd1,lvd2}. Since there are several factors that affect
the calculation of muon intensity underground (muon cross-sections,
muon energy spectrum at surface, slant depth distribution and rock
composition), comparison between measured and calculated muon
intensities does not provide an accurate test of the muon transport code.
In fact the energy spectrum of muons at surface has been reconstructed
from the measured intensities assuming that other factors are known.
However, the fact that measured intensities agree with simulations
over a large range of zenith angles and slant depths, provides a strong evidence
for the validity of the muon transport code.

Depth -- vertical muon intensity relation (muon intensity at vertical
as a function of depth) is shown in Figure \ref{fig-int} for standard rock and water.
Muon intensities have been calculated by integrating muon energy spectra
underground over energy. The simulated curves agree well with the measurements
both in rock (black triangles -- compilation of data points from Ref. \cite{crouch}) and
water (blue open circles -- \cite{baikal}, blue filled circles -- \cite{amanda}).

A comprehensive comparison of calculated (using MUSIC)
muon intensities underground
with measurements has been done in Ref.
\cite{tang}. Data points were found to be scattered symmetrically around the calculated
depth--intensity curve showing the overall consistency of the muon transport.
Large spread of data around simulations may be explained by the complexity
of factors involved in data interpretation, such as, rock composition,
procedure of data conversion to standard rock etc. 

Muon intensities and mean muon energies, 
calculated with the MUSIC transport code and the LVD
parameterisation for the muon spectrum at surface \cite{lvd1,lvd2} (for full description
of the procedure see Section \ref{musun} below)
are given in Table 1 for standard rock and water.

MUSIC has been used in the analysis of SNO \cite{sno} and MACRO \cite{macro}
data. The code has also been applied for the calculation of expected background
induced by cosmic-ray muons in deep underground experiments, such
as KamLAND, Super-Kamiokande, etc.

\begin{figure}[htb]
   \includegraphics[width=15.cm]{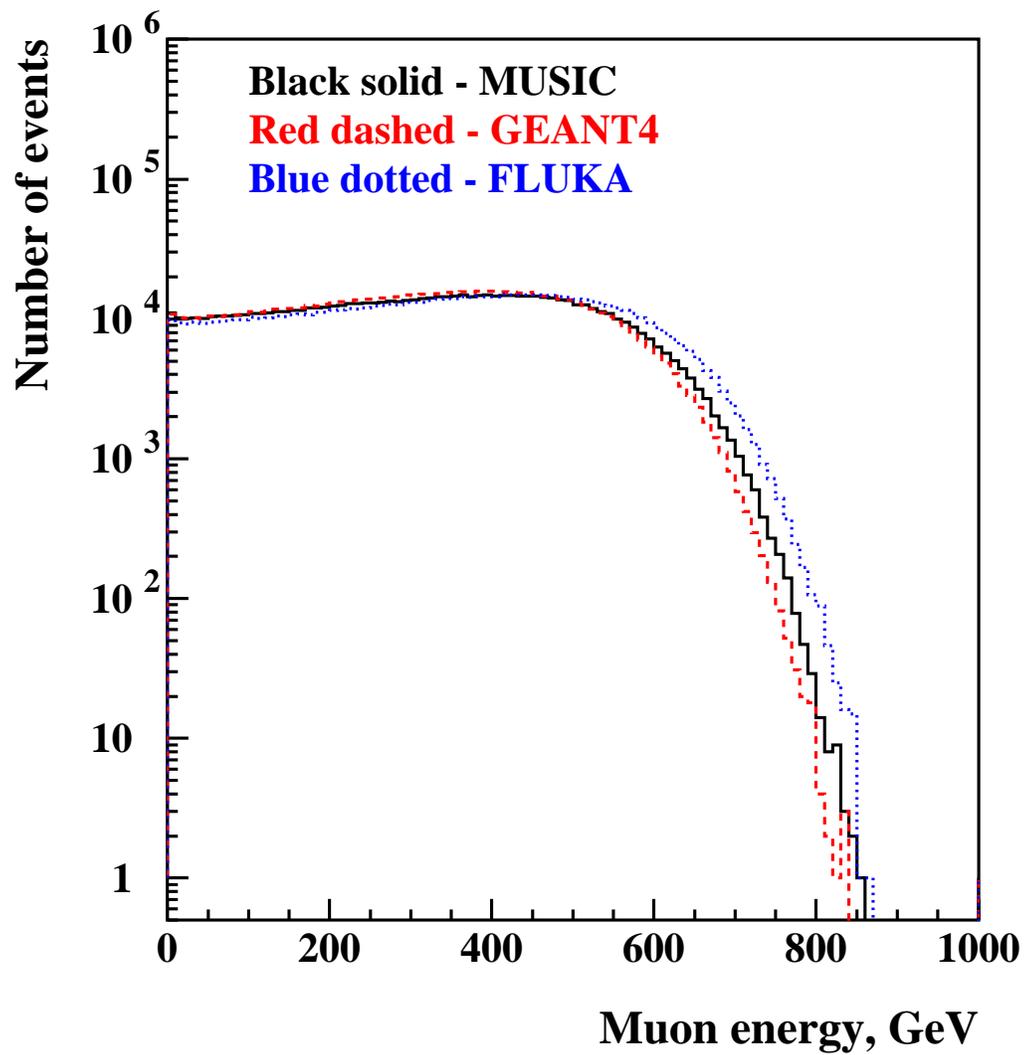}
    \caption{Energy distribution of muons 
with initial energy of 2 TeV transported
through 3 km of water using MUSIC (black solid curve), GEANT4 (red dahsed curve)
and FLUKA (blue dotted curve).}
  \label{fig-energy}
\end{figure}

\begin{figure}[htb]
   \includegraphics[width=15.cm]{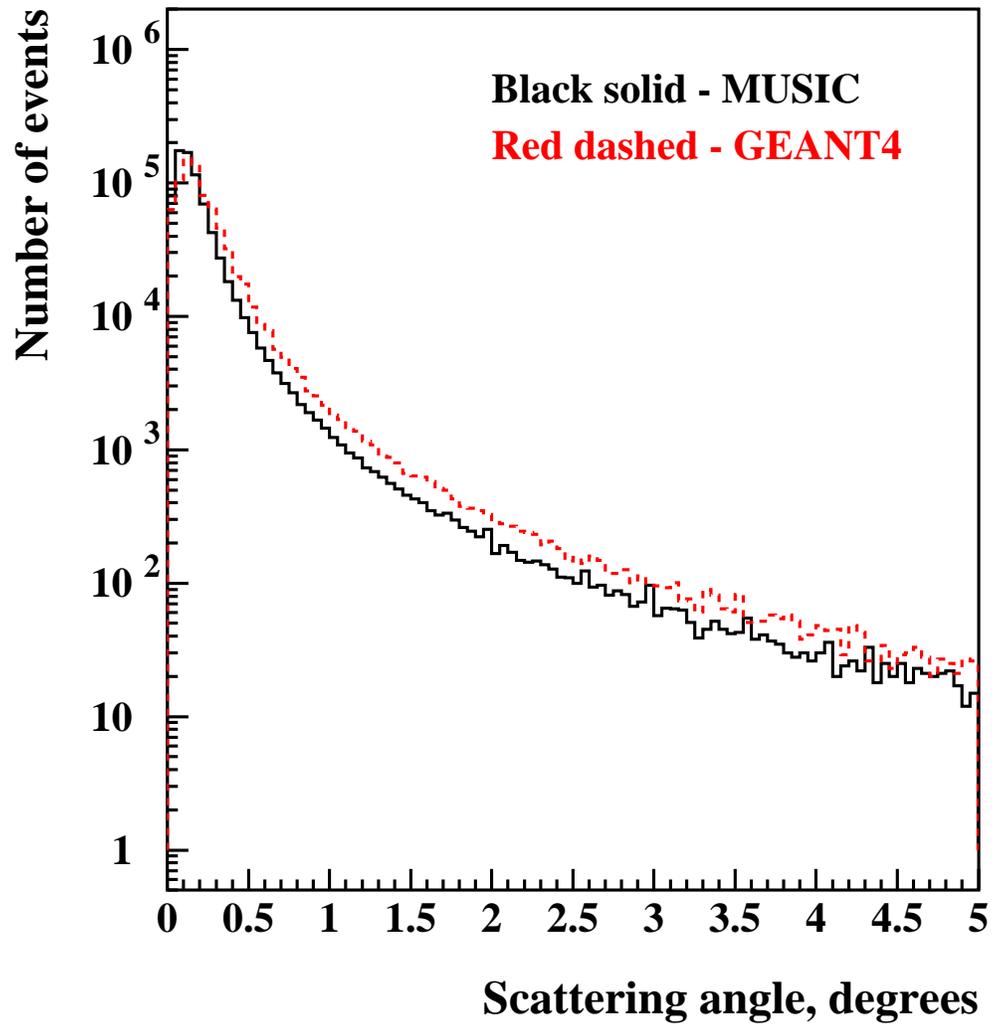}
    \caption{Distribution of angular deviation for muons 
with initial energy of 2 TeV transported
through 3 km of water using MUSIC (black solid curve) and GEANT4 (red dahsed curve).}
  \label{fig-angular}
\end{figure}

\begin{figure}[htb]
   \includegraphics[width=15.cm]{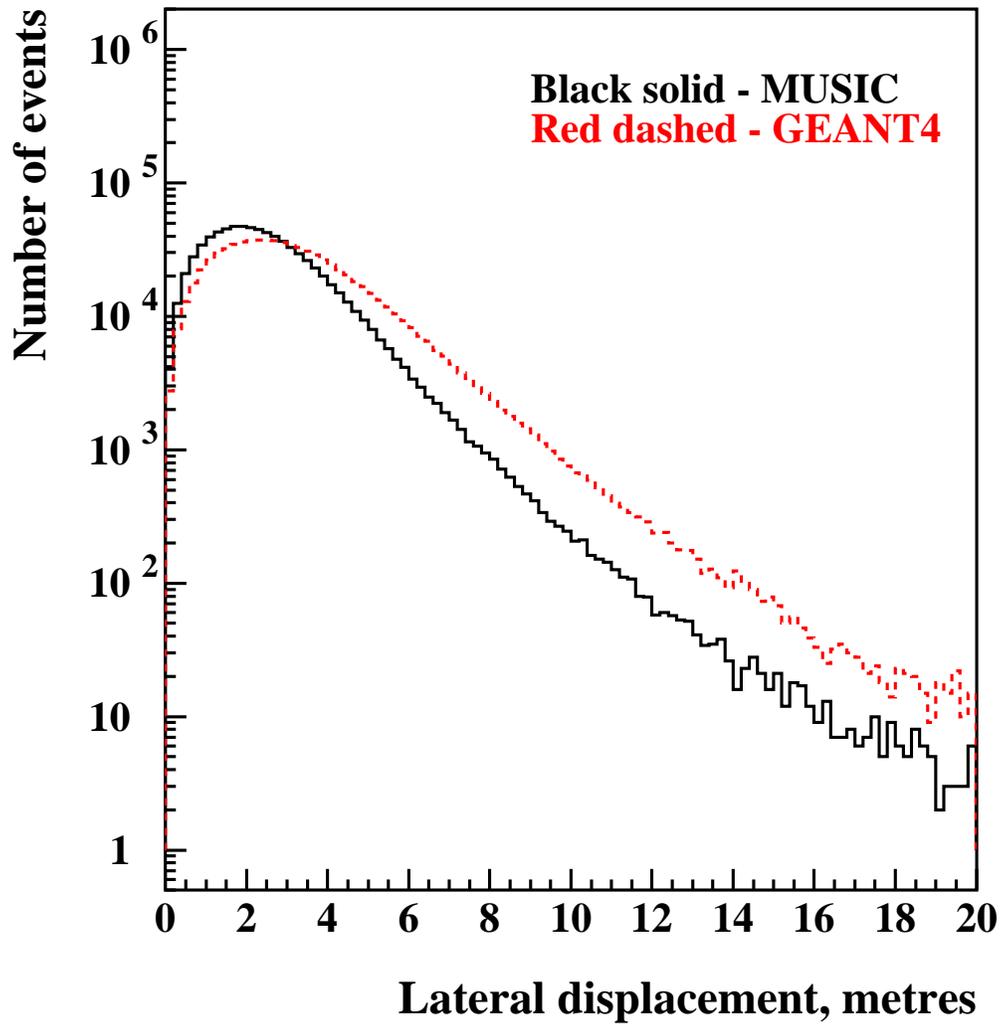}
    \caption{Distribution of lateral displacement for muons 
with initial energy of 2 TeV transported
through 3 km of water using MUSIC (black solid curve) and GEANT4 (red dahsed curve).}
  \label{fig-lateral}
\end{figure}

Comparison of energy losses as calculated by
MUSIC, GEANT4 and FLUKA is discussed in Ref. \cite{gtest}.
All codes agree well in calculating energy distributions for high-energy
muons transported through small and large slabs of materials.
Figure \ref{fig-energy} shows the energy spectrum of muons 
with initial energy of 2 TeV transported
through 3 km of water using MUSIC, GEANT4 \cite{matt} and FLUKA.
Distributions look very similar except for small difference at high energies.
The survival probability is equal to 0.779 (MUSIC), 0.793 (GEANT4) and 0.756 (FLUKA)
with a statistical error of about 0.001.
The mean energy of survived muons is 323 GeV (MUSIC), 317 GeV (GEANT4) and
344 GeV (FLUKA).
Figures \ref{fig-angular} and \ref{fig-lateral} show distributions of
angular deviation and lateral displacement, respectively, 
for muons with initial energy of 2 TeV transported
through 3 km of water. GEANT4 predicts larger
number of muons to be scattered to high angles and moved to large
distances in the plane perpendicular to the muon direction.
The mean scattering angle is equal to $0.22^{\circ}$ (MUSIC) and
$0.27^{\circ}$ (GEANT4), whereas the mean displacement in the
plane perpendicular to the initial muon direction is found to be 2.6 metres
(MUSIC) and 3.3 metres (GEANT4).
Unfortunately it is practically impossible to obtain data on high-energy
muon scattering beyond very large thicknesses of matter to test codes, since 
the lateral separation of muon bundles underground is largely dominated
by the scattering angle of the muon parent in the atmosphere at the
interaction point where this parent is produced.

Similar transport code has also been developed for tau-leptons: TAUSIC
(TAU Simulation Code).

\section {Simulations of muons in underground laboratories using MUSUN}
\label{musun}

MUSUN (MUon Simulations UNderground) is the muon generator useful for
sampling muons in underground laboratories according to their energy
spectrum and angular distribution. It uses the results of muon transport through
matter carried out with MUSIC, convoluted with the muon energy spectrum
and angular distribution at surface. 

At the first stage muons with various initial energies (from 100 GeV to $10^7$~GeV
with a step of $\Delta \log E = 0.025$) are propagated through matter and their energy
distributions at distances from the initial point ranging from 100 m~w.~e. to 15000 m~w.~e.
are written on the computer disk for further processing. This is usually done by the code
developer following instructions from a user about the rock composition and other possible
specific features such as mountain profile etc. In a standard version of MUSUN the
vertical depth should be more than 500 m~w.~e. There is no strict upper limit for the
vertical depth, but the maximum slant depth should not exceed 15 km~w.~e. At larger
depths neutrino-induced muon flux dominates over atmospheric muons and the
calculation of the atmospheric muon intensity is not required.

In a simple version of MUSUN, the flat profile is assumed for the surface above the
underground site (the curvature of the Earth is taken into account but other
possible fluctuations of the slant depth are ignored).

After muon transport the differential muon intensities underground,  
$I_{\mu}(E_{\mu},X,\cos\theta)$, are calculated 
using the equation:

\begin{equation}
I_{\mu}(E_{\mu},X,\cos\theta)=\int_0^{\infty} P(E_{\mu},X,E_{\mu 0})
\frac {{d I_{\mu 0}(E_{\mu 0},\cos\theta^{\star})}} {d E_{\mu 0}}
d E_{\mu 0}
\label{muon intensity}
\end{equation}

\noindent where 
$\frac {{d I_{\mu 0}(E_{\mu 0},\cos\theta^{\star})}} {d E_{\mu 0}}$
is the muon spectrum at sea level at zenith angle
$\theta^{\star}$ (zenith angle at surface, $\theta^{\star}$,
is calculated from the zenith angle
underground, $\theta$, taking into account the curvature of the Earth), and
$P(E_{\mu},X,E_{\mu 0})$ is the probability for a muon with an initial
energy at surface $E_{\mu 0}$ to have an energy $E_{\mu}$ at a
depth $X$.

The energy spectrum at sea level can be taken either according to the 
parameterisation proposed by Gaisser \cite{gaisser} (modified 
for large zenith angles \cite{lvd1}) or following the best fit to the 
`depth -- vertical muon intensity' relation measured by the LVD 
experiment \cite{lvd1}. The first parameterisation \cite{gaisser} has 
the power index of the primary all-nucleon spectrum 2.70, while the 
second one \cite{lvd1} uses the index of 2.77 with the normalisation 
to the absolute flux measured by LVD. 
For small depths (less than 2--3 km~w.~e.) that correspond to low muon 
energies at surface (less than 1 TeV) it is recommended to use
the original Gaisser's parameterisation with an additional factor that
takes into account muon decay in the atmosphere \cite{dar}
if necessary. For larger depths the LVD parametrisation is the preferred
option since it agrees with experimental data of the LVD \cite{lvd1} and MACRO
experiments \cite{macro1}.

The ratio of prompt muons (from charmed particle decay) 
to pions is recommended to be set to
$10^{-4}$, which is well below an upper limit set by the LVD 
experiment \cite{lvd2}. Note, however, that prompt muon flux does 
not affect much muon intensities even at large depths.

To calculate integral muon intensity for normalisation, an integration of 
$I_{\mu}(E_{\mu},X,\cos\theta)$ over $dE_{\mu}$ and $\cos\theta$ is carried out. 

MUSUN offers the choice of the muon energy spectrum (as described above),
the fraction of prompt muons, the vertical depth of the laboratory, the range of
zenith and azimuthal angles, and the range of energies.
No additional muon propagation is required for different options.
Different types of rocks (rock compositions), however, require separate
muon transport.

MUSUN is organised as a set of subroutines written in FORTRAN that are
called from the user-defined 'main' program. The first call is made to a subroutine
that calculates differential and integrated muon intensities for a specific vertical
depth (assuming flat surface). The intensity as a function of energy and
zenith angle is stored in the computer
memory as a two-dimensional array. Subsequent calls to a 'sampling'
subroutine return muon parameters (energy and direction cosines)
sampled following energy and zenith angle distribution. Azimuthal
angle is sampled randomly as evenly distributed between 0 and 2$\pi$
since the assumption of the flat
surface leads to the spherical symmetry. The muon charge is
generated according to the ratio measured for high-energy muons
$\mu^{+}/\mu^{-} \approx 1.3$.

For practical purposes (for instance, when these muons are used
in multipurpose event generators GEANT4 or FLUKA)
it is useful to generate muons on the surface of a rectangular
parallelepiped or a sphere with predefined dimensions.
MUSUN offers a possibility to generates muon positions on the
surface of a rectangular parallelepiped with dimensions specified by 
the user. 

Muon parameters are written on the disk and can be passed later on 
to the multipurpose event generators.

Several underground laboratories (for instance, LNGS at Gran Sasso 
and LSM at Modane)
are located in the transport tunnels under mountains with complex
mountain profiles. For these labs special versions of the MUSUN code
have been developed that took into account the slant depth
distribution as seen from the underground laboratory.
Here we present a few graphs with the results of muon production using MUSUN
for the Gran Sasso Laboratory. 

\begin{figure}[htb]
   \includegraphics[width=15.cm]{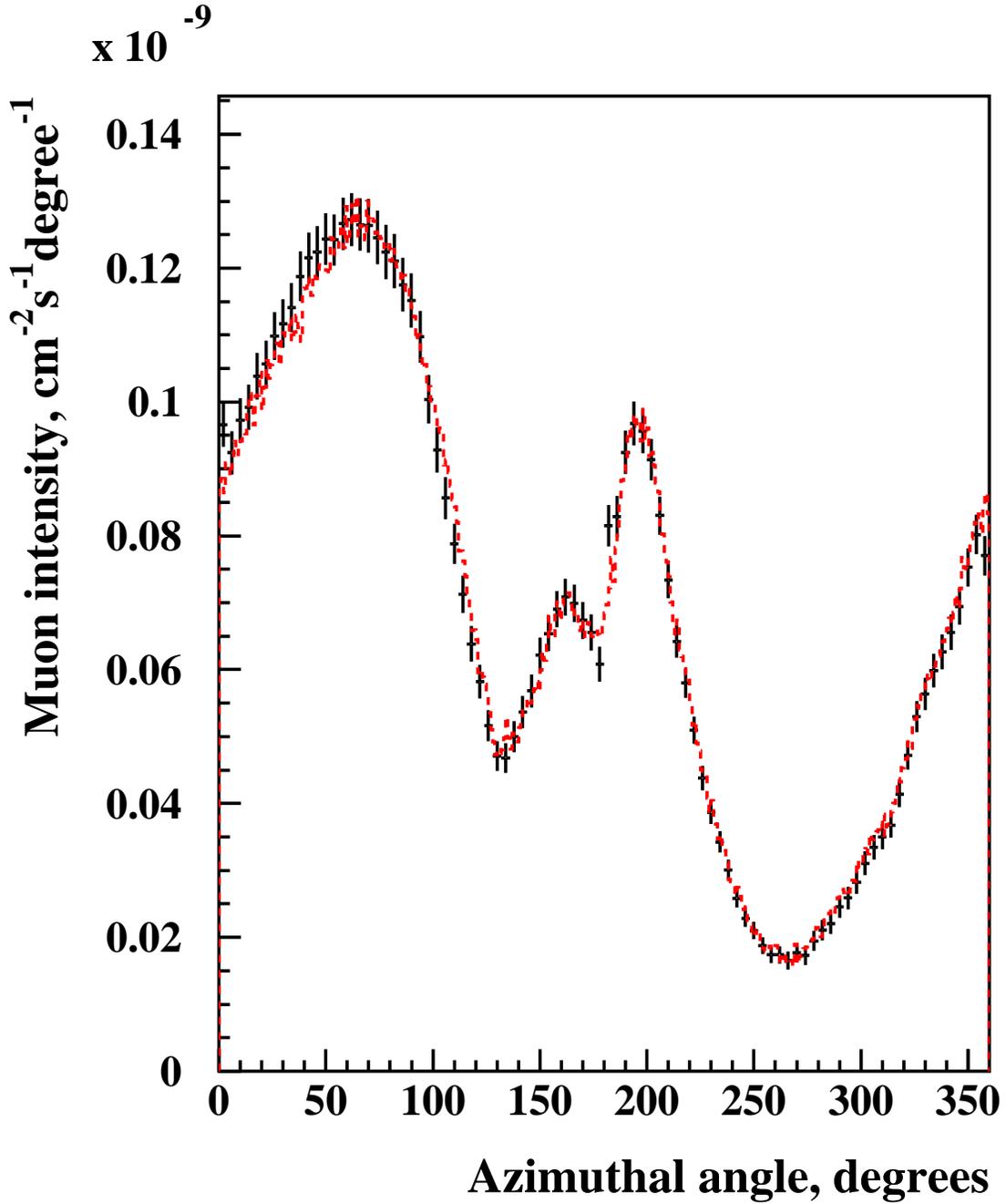}
    \caption{Azimuthal distribution of single muon intensities
in the underground Gran Sasso Laboratory for zenith angles up to $60^{\circ}$
as measured by LVD \cite{lvd2,lvd3} (data points with error bars)
and generated with MUSUN (dashed curve). LVD acceptance as a function of zenith
and azimuthal angles has been taken into account when generating muons.
Azimuthal angle is calculated in the LVD
reference system \cite{lvd1,lvd2,lvd3}.}
  \label{fig-azimuth}
\end{figure}

Figure \ref{fig-azimuth} shows the azimuthal distribution of single muon intensities
in the underground Gran Sasso Laboratory for zenith angles up to $60^{\circ}$
as measured by LVD \cite{lvd2,lvd3} (data points with error bars)
and generated with MUSUN (dashed curve). Good agreement is seen over the whole
range of angles and intensities. Similar conclusion has been achieved in Ref.
\cite{pandola} when comparing azimuthal distributions for the whole  range of 
zenith angles.

\begin{figure}[htb]
   \includegraphics[width=15.cm]{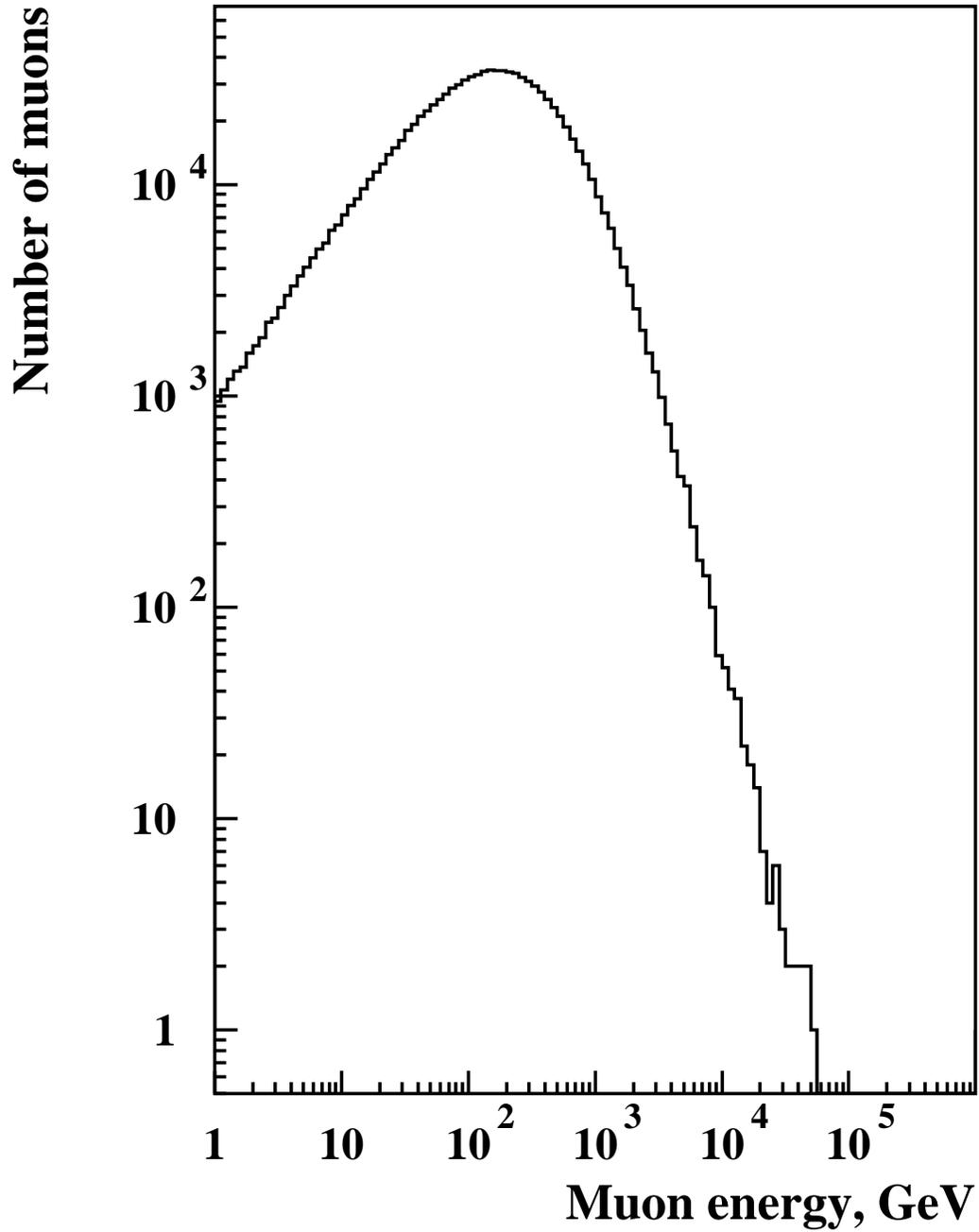}
    \caption{Energy spectrum of muons as generated by MUSUN
for the underground Gran Sasso Laboratory.}
  \label{fig-gs-energy}
\end{figure}

\begin{figure}[htb]
   \includegraphics[width=15.cm]{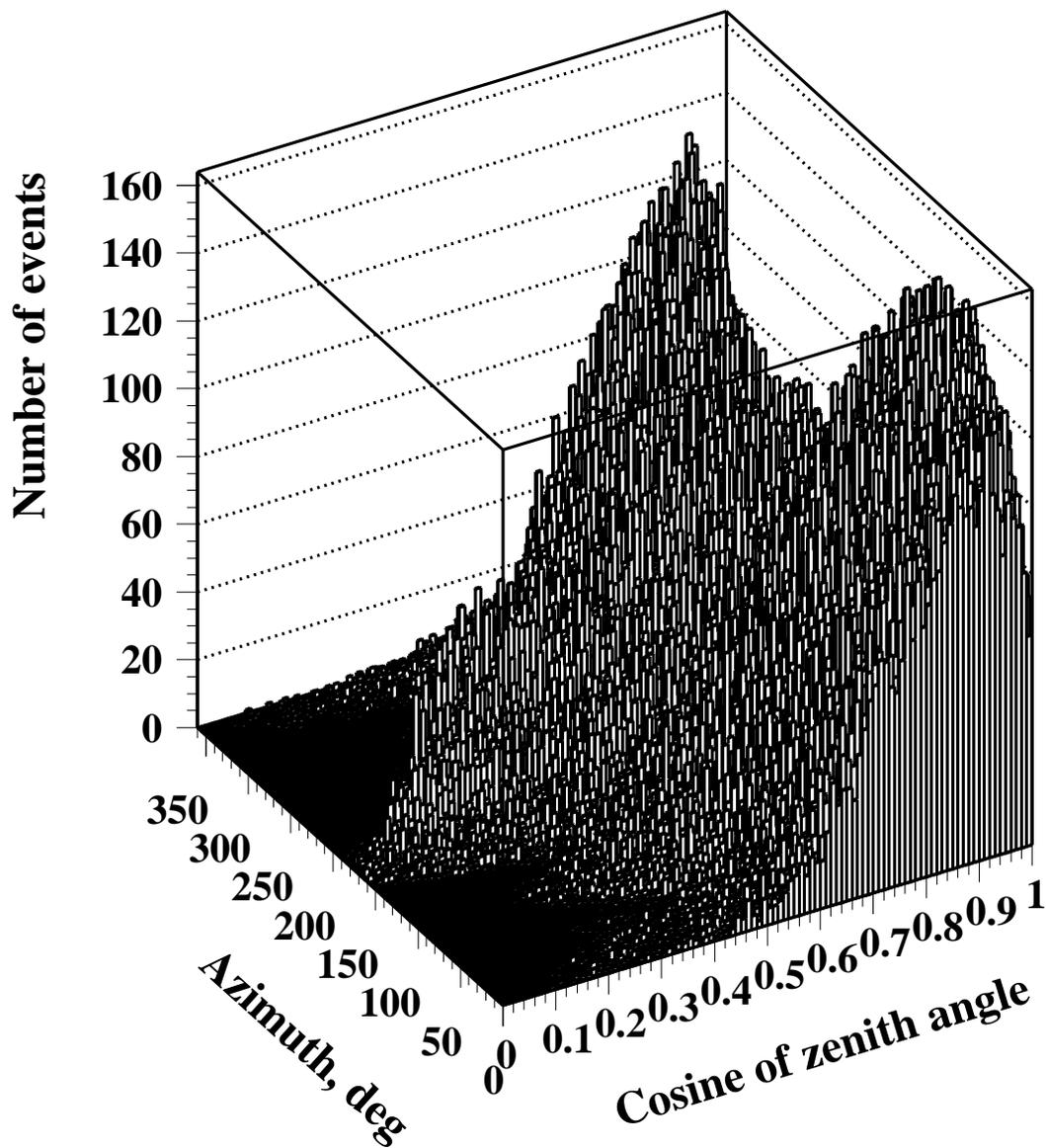}
    \caption{A lego plot of the number of generated muons (using MUSUN)
as a function of zenith and azimuthal angles in the Hall A of the
underground Gran Sasso Laboratory. Azimuthal angle is calculated in the LVD
reference system \cite{lvd1,lvd2,lvd3}.}
  \label{fig-lego}
\end{figure}

Figure \ref{fig-gs-energy} shows the energy spectrum of muons at Gran Sasso as 
generated with MUSUN. This spectrum looks different from Figure \ref{fig-musp-energy} 
because the number of muons is given here per energy bin which is constant on the logarithmic
scale but increases with energy on the linear scale, whereas in Figure \ref{fig-musp-energy}
the spectrum is given per constant energy bin on the linear scale 
(1 GeV). The mean muon energy at the
Gran Sasso Laboratory is calculated as 273 GeV, in good agreement with the measured
value of $270\pm3$ (stat.) $\pm18$ (syst.) GeV \cite{macro-energy}.
A lego plot of the number of generated muons versus
zenith and azimuthal angles is presented in Figure \ref{fig-lego}.

At present the versions of the MUSUN code exist for the underground sites
at Gran Sasso (LNGS), Modane (LSM), Boulby and Soudan.
It has been used to study muon-induced neutron background for experiments
looking for rare events, such as WIMPs (see, for instance,
\cite{musun,carson04,carson05,araujo05,wulandari04,vak08}).

\section {Conclusions}
\label{conclusions}
The two Monte Carlo codes MUSIC and MUSUN dedicated to muon simulations
have been described. MUSIC, a package for muon transport through matter,
can be used for propagating
muons through large thickness of rock or water, for instance from the surface down to
underground/underwater laboratory. It can also be implemented
in the event generators for large underwater/under-ice neutrino telescopes or other
neutrino detectors.
MUSUN uses the results of muon
transport through rock/water to generate muons in or around underground laboratory
taking into account their energy spectrum and angular distribution.
Various tests showed good agreement of the codes' results
with experimental data and other packages.

Since there are several versions of both codes the author finds 
impractical to submit all of them to the
code library. Any specific version can be obtained by request to 
v.kudryavtsev@sheffield.ac.uk.
There is a possibility to adapt the codes to specific needs of a user as was done on several
occasions in the past.

\section{Acknowledgments}
Many recent improvements to the codes have been 
carried out as part of the ILIAS integrating activity 
(Contract No. RII3-CT-2004-506222) in the framework of the EU FP6 programme 
in Astroparticle Physics.
The author is grateful to Dr. M. Robinson for providing results of GEANT4
simulations. The author wishes to thank Drs. P.~Antonioli, C.~Ghetti, 
E.V.~Korolkova and G.~Sartorelli who contributed to the original version of the MUSIC
code and initial tests.

\pagebreak

\begin{table}[htb]
\caption{Muon intensities and mean energies at various depths underground for standard 
rock and water calculated using the MUSIC code for muon transport
and the LVD parameterisation for the muon spectrum at surface \cite{lvd1,lvd2}.
Flat surface relief was assumed but the curvature of the Earth was taken into account.
Column 1 -- depth, $X$, in kilometres of water equivalent, km~w.~e.; 
column 2 -- vertical muon intensity in standard rock, $I_{\mu}^{vert}$; 
column 3 -- mean muon energy for the muon flux in standard rock at vertical, $E_{\mu}^{vert}$;
column 4 -- global intensity (integrated 
over solid angle for a spherical detector) in standard rock for flat surface, $I_{\mu}$; 
column 5 -- mean muon energy for the global muon flux in standard rock, $E_{\mu}$;
column 6 -- vertical muon intensity in water (ice); 
column 7 -- mean muon energy for the muon flux in water at vertical;
column 8 -- global intensity (integrated 
over solid angle for a spherical detector) in water (ice) for flat surface; 
column 9 -- mean muon energy for the global muon flux in water.}
\label{intensity}
\vspace{1cm}
\begin{footnotesize}
\begin{center}
\begin{tabular}{|c|c|c|c|c|c|c|c|c|}\hline
& \multicolumn {4} {c|} {Standard rock} & \multicolumn {4} {c|} {Water} \\ 
\hline
$X$& $I_{\mu}^{vert}$ & $E_{\mu}^{vert}$ & $I_{\mu}$ & $E_{\mu}$ & 
$I_{\mu}^{vert}$ & $E_{\mu}^{vert}$ & $I_{\mu}$ & $E_{\mu}$ \\
km w.e. & cm$^{-2}$s$^{-1}$sr$^{-1}$ & GeV & cm$^{-2}$s$^{-1}$ & GeV &
cm$^{-2}$s$^{-1}$sr$^{-1}$ & GeV & cm$^{-2}$s$^{-1}$ & GeV \\
\hline
0.5 & $1.06 \times 10^{-5}$  &   69 & $2.07 \times 10^{-5}$ &   93 &
          $7.86 \times 10^{-6}$  &   80 & $1.58 \times 10^{-5}$ &  111 \\
1.0 & $1.47 \times 10^{-6}$  & 120 & $2.56 \times 10^{-6}$ & 150 &
          $1.14 \times 10^{-6}$  & 144 & $2.09 \times 10^{-6}$ & 186 \\
2.0 & $1.38 \times 10^{-7}$  & 197 & $1.99 \times 10^{-7}$ & 225 &
          $1.22 \times 10^{-7}$  & 246 & $1.90 \times 10^{-7}$ & 291 \\
3.0 & $2.56 \times 10^{-8}$  & 248 & $3.09 \times 10^{-8}$ & 271 &
          $2.61 \times 10^{-8}$  & 322 & $3.53 \times 10^{-8}$ & 362 \\
4.0 & $6.16 \times 10^{-9}$  & 284 & $6.38 \times 10^{-9}$ & 301 &
          $7.40 \times 10^{-9}$  & 379 & $8.81 \times 10^{-9}$ & 412 \\
5.0 & $1.70 \times 10^{-9}$  & 308 & $1.53 \times 10^{-9}$ & 319 &
          $2.44 \times 10^{-9}$  & 421 & $2.58 \times 10^{-9}$ & 447 \\
6.0 & $5.06 \times 10^{-10}$ & 324 & $4.02 \times 10^{-10}$ & 332 &
          $8.78 \times 10^{-10}$ & 453 & $8.32 \times 10^{-10}$ & 474 \\
7.0 & $1.58 \times 10^{-10}$ & 335 & $1.12 \times 10^{-10}$ & 341 &
          $3.35 \times 10^{-10}$ & 477 & $2.87 \times 10^{-10}$ & 492 \\
8.0 & $5.07 \times 10^{-11}$ & 344 & $3.23 \times 10^{-11}$ & 347 &
          $1.33 \times 10^{-10}$ & 495 & $1.03 \times 10^{-10}$ & 506 \\
9.0 & $1.67 \times 10^{-11}$ & 349 & $9.61 \times 10^{-12}$ & 351 &
          $5.41 \times 10^{-11}$ & 508 & $3.84 \times 10^{-11}$ & 516 \\
10.0& $5.55 \times 10^{-12}$ & 351 & $2.91 \times 10^{-12}$ & 353 &
            $2.24 \times 10^{-11}$ & 519 & $1.46 \times 10^{-11}$ & 524 \\
\hline
\end{tabular}
\end{center}
\end{footnotesize}
\end{table}

\pagebreak

\pagebreak

\pagebreak

\pagebreak

\pagebreak

\pagebreak

\pagebreak

\pagebreak

\pagebreak

\pagebreak

\end {document}